# An Optical Fibre Pulse Rate Multiplier for Ultra-low Phase-noise Signal Generation

A. Haboucha,[1] W. Zhang,[1] T. Li,[1,2] M Lours,[1] A. N. Luiten,[3] Y. Le Coq,[1] and G. Santarelli[1,]

[1]*LNE-SYRTE, Observatoire de Paris, CNRS, UPMC, 61 Avenue de l'Observatoire, 75014 Paris, France*
[2]*Key Laboratory of Quantum optics, Shanghai Institute of Optics and Fine Mechanics, CAS, 201800 Shanghai, PRC*
[3]*School of Physics, University of Western Australia, Crawley 6009, Australia*

In this letter we report on an all optical-fibre approach to the synthesis of ultra-low noise microwave signals by photodetection of femtosecond laser pulses. We use a cascade of Mach-Zehnder fibre interferometers to realize stable and efficient repetition rate multiplication. This technique increases the signal level of the photodetected microwave signal by close to 18 dB, that in turn, allows us to demonstrate a residual phase noise level of -118 dBc/Hz at 1 Hz and -160 dBc/Hz at 10 MHz from 12 GHz signal. The residual noise floor of the fiber multiplier and photodetection system alone is around -164 dBc/Hz at the same offset frequency, which is very close to the fundamental shot noise floor.

Low-phase-noise and frequency stable microwave signals are crucial in a wide variety of scientific and technological applications including precise timing, phased-array radars, arbitrary waveform generation, photonic processing and atomic frequency standards [1-5]. At present, the lowest noise microwave sources are based on ultra-low noise sapphire, or optoelectronic oscillators [6,7]. Despite the superb performance of these existing devices, there continues to be great interest in developing simpler or more robust devices that can equal or better their performance. In particular one challenging aim for researchers has been the development of a single device that exhibits low phase fluctuations across the spectrum from low Fourier frequencies (1 Hz) out to the highest (>1 MHz). One possible route to achieving such comprehensive performance would be low-noise frequency division of a laser that has been stabilized to a mode of a vibration-insensitive reference cavity [8,9]. In this approach, the divided signal would carry the frequency stability of the original while the division process itself improves the signal-to-noise ratio by the division ratio. In this letter we demonstrate low noise frequency division using a robust combination of a fiber-based mode-locked laser and a cascaded set of fiber interereferometers.

There have been several recent demonstrations of the use of femtosecond frequency combs to generate very low absolute or residual phase noise microwave signals [10,11]. These combs have been generated using either mode-locked Ti:Sapphire and Erbium fiber lasers, and while the fibre systems are more compact, robust and power efficient, they suffer a lower pulse repetition rate frequency, which reduces the available power at a given harmonic of the repetition rate. For instance, a standard photodetector receiving the output of a commercial 250 MHz repetition rate frequency comb will saturate at an incident optical power of just 1 mW (4 pJ per pulse) at which point the output power of a 10 GHz harmonic signal will be around -30 dBm. Although research is being pursued on higher linearity photodetectors (e.g. we have obtained 10 dB improvement in the level of this harmonic using a Highly Linear Photo Diode, (Discovery Semiconductor, HLPD 40 S). The low power on a given harmonic precludes the possibility to generate signals with phase noise much better than -140 dBc/Hz (set by the comparison of signal power and the thermal and shot noise which dominate sufficiently far from the carrier). Recently, using custom-designed photodiodes, the NIST group have demonstrated an improved phase noise floor of -145 dBc/Hz [12].

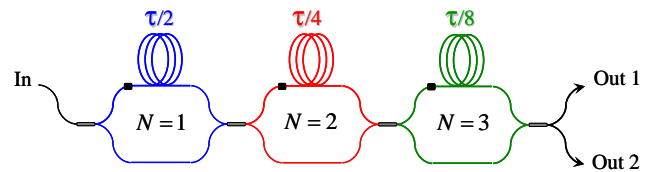

Figure 1 An illustration of the cascaded Mach-Zehnder interferometer (MZI) scheme used to achieve a pulse rate multiplication.

Two approaches have been suggested to circumvent this low pulse rate limitation of mode-locked fibre lasers. The first approach makes use of optical spectrum filtering with a Fabry-Pérot cavity [13]. This method, requiring both fine alignment and a frequency lock system, suffers an intrinsic power loss proportional to the desired multiplication factor. This power loss exactly cancels the benefits that accrue from the pulse rate multiplication thereby limiting the potential for a high signal to noise signal. The losses can be, in principle, compensated for by post-cavity optical amplification, but this can lead to excess phase and amplitude noise as well as Amplified Spontaneous Emission noise (ASE) from the amplifier. An alternative technique to realize pulse repetition rate multiplication (PRRM) is to use fiber interferometers. This approach has already being successfully used in optical signal processing, optical communications, photonic signal processing and mode locked laser pulse rate multiplication [14,15]. We implement cascaded Mach-Zehnder interferometers (MZI) for the PRRM as shown in Fig 1. This configuration realizes a $2^N$ repetition rate multiplication where *N* is the number of stages. The devices are realized using off-the-shelf 2x2 single mode fiber couplers. We hand-select devices which have a coupling ratios within 1% of the nominal 50% together with a low insertion loss (<0.3 dB). The topology is realized by fusion spliced connections (<0.05 dB loss each).

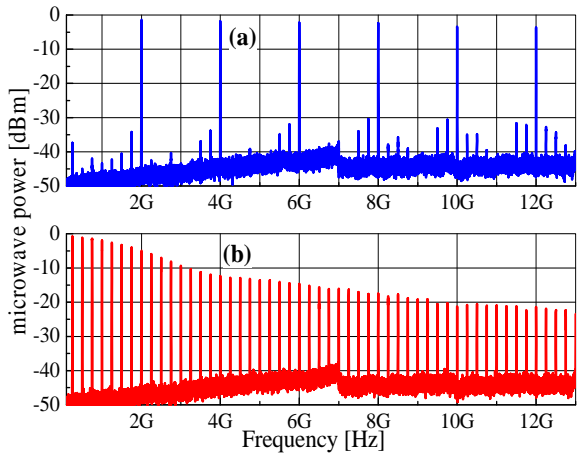

Figure 2 Photodiode (HLPD) output driven with 10 mW optical power. Plot (a) after PRRM x8. Plot(b) driven by the unmultiplied FOFC output.

These precautions ensure a low amplitude mismatch of the MZIs. The path length difference between the two arms is adjusted to obtain a propagating delay equal to half the inter-pulse duration of the input pulses to each individual stage. With simple fiber length measurements, the delay difference can be coarsely adjusted to within approximately a percent of the desired value. Additional fine tuning is achieved by varying the repetition rate of the input pulse train and measuring the resulting spectrum of the output signal from a photodetector coupled to the output of the MZI. The arms are then respliced to maximize the rejection of the odd harmonics of the repetition rate when the FOFC is tuned to the nominal frequency of 250 MHz. Using this iterative procedure we approach the optimal path length difference within a few attempts. We have constructed two three-stage PRRMs exhibiting extremely good performance: the total insertion loss, between the input and one of the output ports, for the three-stage devices was 3.5 dB and 3.8 dB respectively. One notes that the intrinsic insertion loss of this topology is 3 dB due to the splitting of the power between the two output ports. In principle, one can recover the 3 dB by using photodiodes on both output ports and coherently combining their outputs electrically. Fig 2 (a) shows the typical microwave spectrum revealed by a fast photodiode at the output of a MZI PRRM. The low levels of unwanted harmonics across the whole spectrum demonstrate the accuracy of the length and amplitude matching of the individual MZI devices. Fig. 2 b shows the typical output spectrum of a saturated HLPD driven directly by the comb without the benefit of the PRRM. The power level gain for the 12 GHz signal is about 17 dB, which is very close to the expected 18 dB.

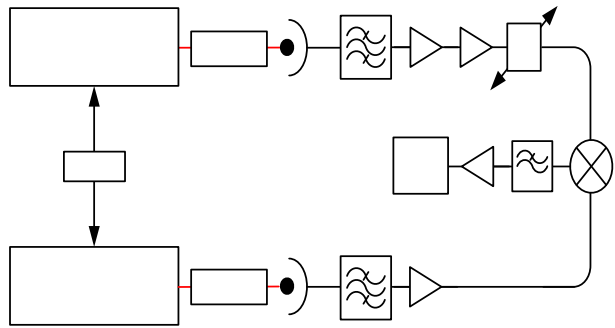

Figure 3 Experimental setup. USL: ultra-stable laser; PRRM: Pulse Repetition Rate Multiplier; HLPD: Highly Linear Photodiode, FFT: Fast Fourier Transform analyzer.

Another issue, which becomes critical at the low phase noise levels expected from our PRRM approach, is a complex nonlinear amplitude to phase conversion process arising in the photodetection process [16,17]. Very briefly, as shown in [17], for some specific combinations of bias voltage and optical energy per pulse, the static coefficient which relates the amplitude fluctuations of the laser to the phase fluctuations of the microwave signal can be zeroed. We can reliably operate near one of these null points, which ensures a very low amplitude-to-phase conversion [17]. In these conditions the impact of the FOFC's RIN is below -165 dBc/Hz beyond 1 kHz. All measurements presented here were performed in this regime.

We have made two different measurements: the first to quantify the additive phase noise of the fibre pulse rate multiplier, while in the second we estimate the additive phase noise of the fibre comb combined with the pulse rate multiplier. For the first measurement we drive two identical three-stage PRRMs with the same FOFC. The differential phase noise is measured by isolating the 12 GHz harmonics using narrow bandpass filters, amplifying these signals with ultra low phase noise microwave amplifiers (-130 dBc/Hz@1Hz, -170 dBc/Hz beyond 100 kHz, and then comparing their phase with a well characterized double balanced mixer. The mixer output is amplified and sampled by a 10 MHz FFT analyzer. Fig 4a shows the measured phase noise from this measurement, which is solely associated with the PRRMs and detection since the FOFC and optical reference is common to both signals. The measurement shows the beneficial effect of the PRRM and the HLPD. We simultaneously achieve -118 dBc/Hz at 1 Hz (one system) and a floor approaching -164 dBc/Hz beyond 1 MHz showing the excellent potential for this approach to transfer the full frequency stability of the input pulse train while also preserving the excellent signal-to-noise ratio of the input signal. This noise floor is mainly determined by the photocurrent shot-noise (dotted line in Fig. 4) [18,19]. In the second measurement we measure the phase noise between two PRRMs driven by two independent FOFCs that are phase-locked to a common high performance optical frequency reference [10,20].

The result is shown in Fig. 4b. When compared to the earlier measurements (curve a on Fig 4), this measurement includes the additional noise associated with the FOFCs and their lock to the optical reference. We can interpret this result as an estimation of the potential to transfer the phase stability of the optical source into the microwave domain. Curves a and b on Fig 4 are very close together below Fourier frequencies of 100 Hz showing that the FOFC lock is not a limiting factor in this range.

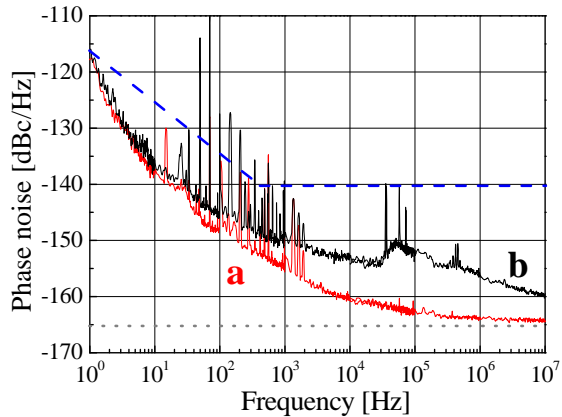

Figure 4 Plot (a) Residual phase noise for a single PRRM system (measured with a common FOFC). The dotted line estimates the shot-noise limit for these conditions (~11 mW optical power, 8 mA photocurrent). Plot (b) Residual phase noise for a single complete optical-to-microwave system i.e. an FOFC followed by a PRRM and HPLD (measured with two independent FOFCs). The dashed line represents the typical phase noise level that would be achievable without the PRRM.

At higher Fourier frequencies, the two curves differ substantially. This extra noise is partially explained by imperfect phase lock loops in the 1 kHz to 1 MHz region. Beyond 1 MHz the excess noise level could not be explained by the free running optical frequency noise of the FOFCs. Moreover the noise level and shape depended upon the mode-locking state of the FOFC. Nevertheless we reach a noise level of -160 dBc/Hz at 10 MHz. For comparison the dashed line on Fig. 4 show typical previous results without PRRM and HPLD. To conclude, we have demonstrated microwave signal generation with very low residual phase noise using a combination of fiber based frequency combs and fibre-based pulse rate multiplication. This result paves the way to compact, robust and efficient low phase photonic microwave signal generation.

**References.**

1. J. Kim and F. X. Kärtner, Opt. Lett. 35, 2022-2024 (2010) and references therein.
2. S. A. Diddams, J. Opt. Soc. Am. B 27, B51-B62 (2010) and references therein.
3. N. R. Newbury, Nat Phot., 5, 186–188 (2011) and references therein.
4. S. Weyers, B. Lipphardt, and H. Schnatz, Phys. Rev. A, 79, 031803R (2009).
5. J. Millo, M. Abgrall, M. Lours, E. M. L. English, H. Jiang, J. Guéna, A. Clairon, S. Bize, Y. Le Coq and G. Santarelli, Appl. Phys. Lett, 94, 141105 (2009).
6. E.N. Ivanov, M.E. Tobar, IEEE Trans. on Ultrason. Ferr. Freq. Contr., 56, 263-269, (2009).
7. D. Eliyahu, D. Seidel, L. Maleki, in Proceedings of the 2008 International Frequency Control Symposium (IEEE, 2008), pp.811-814.
8. J. Millo, D. V. Magalhaes, C. Mandache, Y. Le Coq, E. M. L. English, P. G. Westergaard, J. Lodewyck, S. Bize, P. Lemonde and G. Santarelli, Phys. Rev. A 79, 053829 (2009) and references therein.
9. D. R. Leibrandt, M. J. Thorpe, M. Notcutt, R. Drullinger, T. Rosenband, and J. C. Bergquist, Opt. Express, 19, 3471-3482 (2011).
10. W. Zhang, Z. Xu, M. Lours, R. Boudot, Y. Kersalé, G. Santarelli and Y. Le Coq, Appl. Phys. Lett., 96, 211105 (2010).
11. T. M. Fortier, M. S. Kirchner, F. Quinlan, J. Taylor, J. C. Bergquist, T. Rosenband, N. Lemke, A. Ludlow, Y. Jiang, C. W. Oates and S. A. Diddams, Nature Photonics, 5, 425–429, (2011) and references therein.
12 F. Quinlan, T. M. Fortier, S. Kirchner, J. A. Taylor, M. J. Thorpe, N. Lemke, A. D. Ludlow, Y. Jiang, C. W. Oates, and S. A. Diddams, arXiv:1105.1434v1, in press Opt. Lett. (2011).
13. S. A. Diddams, M. Kirchner, T. Fortier, D. Braje, A. M. Weiner, and L. Hollberg, Opt Express. 17, 3331-3340 (2009).
14. B. Xia and L. R. Chen, IEEE Journal of Selec. Top. in Quant. Elec, 11, (2005) and references therein
15. S. Min, Y. Zhao and S.Fleming, Optics Comm., 277, 411-413, (2007).
16. J.A. Taylor, S. Datta, A. Hati, C. Nelson F. Quinlan, A. Joshi, S.A. Diddams IEEE Photonics Journal, 3, 140-151, (2011)
17. W. Zhang, T Li, M. Lours, S. Seidelin, G. Santarelli, and Y. Le Coq, in press Applied Physics B arXiv:1104.4495v1, (2011).
18 The shot noise is estimated using the classical formula applied to SSB phase noise and corrected for losses and HLPD response (see also P. J. Winzer, J. Opt. Soc. Am. B 14, 2424-2429 (1997)).
19 R.P. Scott, C. Langrock, B.H Kolner, IEEE Journal of Selected Topics in Quantum Electronics, 7, 641-655, (2001).
20 W. Zhang, Z. Xu, M. Lours, R. Boudot, Y. Kersalé, A. N. Luiten, Y. Le Coq, G. Santarelli, IEEE Trans. on Ultrason. Ferr. Freq. Contr., 58, 886-889 (2011).